\documentclass[%
 reprint,
 superscriptaddress,
 showpacs,preprintnumbers,
 amsmath,amssymb,
 aps,
 prb,
]{revtex4-1}

\usepackage{graphicx}
\usepackage{dcolumn}
\usepackage{bm}

\usepackage{upgreek}
\usepackage{amsmath}

\begin{document}


\title{Controlling light absorption of graphene at critical coupling through magnetic dipole quasi-bound states in the continuum resonance}

\author{Xing Wang}
\affiliation{Institute for Advanced Study, Nanchang University, Nanchang 330031, China}
\affiliation{Jiangxi Key Laboratory for Microscale Interdisciplinary Study, Nanchang University, Nanchang 330031, China}

\author{Junyi Duan}
\affiliation{Institute for Advanced Study, Nanchang University, Nanchang 330031, China}
\affiliation{Jiangxi Key Laboratory for Microscale Interdisciplinary Study, Nanchang University, Nanchang 330031, China}

\author{Wenya Chen}
\affiliation{Institute for Advanced Study, Nanchang University, Nanchang 330031, China}
\affiliation{Jiangxi Key Laboratory for Microscale Interdisciplinary Study, Nanchang University, Nanchang 330031, China}

\author{Chaobiao Zhou}
\affiliation{College of Mechanical and Electronic Engineering, Guizhou Minzu University, Guiyang 550025, China}

\author{Tingting Liu}
\email{ttliu@hue.edu.cn}
\affiliation{School of Physics and Electronics Information, Hubei University of Education, Wuhan 430205, China}

\author{Shuyuan Xiao}
\email{syxiao@ncu.edu.cn}
\affiliation{Institute for Advanced Study, Nanchang University, Nanchang 330031, China}
\affiliation{Jiangxi Key Laboratory for Microscale Interdisciplinary Study, Nanchang University, Nanchang 330031, China}

\begin{abstract}
Enhancing the light-matter interaction in two-dimensional (2D) materials with high-$Q$ resonances in photonic structures has boosted the development of optical and photonic devices. Herein, we intend to build a bridge between the radiation engineering and the bound states in the continuum (BIC), and present a general method to control light absorption at critical coupling through the quasi-BIC resonance. In a single-mode two-port system composed of graphene coupled with silicon nanodisk metasurfaces, the maximum absorption of 0.5 can be achieved when the radiation rate of the magnetic dipole resonance equals to the dissipate loss rate of graphene. Furthermore, the absorption bandwidth can be adjusted more than two orders of magnitude from 0.9 nm to 94 nm by simultaneously changing the asymmetric parameter of metasurfaces, the Fermi level and the layer number of graphene. This work reveals out the essential role of BIC in radiation engineering and provides promising strategies in controlling light absorption of 2D materials for the next-generation optical and photonic devices, e.g., light emitters, detectors, modulators, and sensors.
\end{abstract}

\maketitle


\section{\label{sec1}Introduction}

Metasurfaces have received increasing attention due to their extraordinary characteristics. They are composed of periodic arrays of subwavelength resonators whose arrangement and interaction can modify the properties of electromagnetic waves, such as amplitude, phase, polarization and propagation direction\cite{Zheludev2012, Xiao2020}. In particular, those metasurfaces made of high-index dielectric materials have recently emerged as the essential elements in the field because they show a very high diversity of available functionalities while avoiding the ohmic losses associated with their plasmonic counterparts\cite{Kuznetsov2016, Baranov2017}. The non-radiative states with near-infinite lifetime and perfect energy confinement is allowed. The excitations of guided mode\cite{Mocella2015, Yoon2015, Maksimov2020}, trapped mode\cite{Zhang2013, Cui2018, Xu2019, Zhang2019, Tian2020}, toroidal mode\cite{Kaelberer2010, Savinov2014, Tuz2018, Fan2018, He2018}, anapole mode\cite{Miroshnichenko2015, Yang2018, Tian2019}, and supercavity mode\cite{Kodigala2017, Rybin2017} fundamentally linked to the physics of bound states in the continuum (BIC) are nicely observed in dielectric metasurfaces, which provide an approach to extreme light localization and radiation engineering.

In recent years, enhancing the light absorption of two-dimensional (2D) materials with photonic structures has been demonstrated for high-efficiency optical and photonic devices\cite{Xia2014}. Jessica R. Piper designed the completely controlling the critical coupling within a hybrid graphene/photonic crystal structure, where graphene absorbs up to $100\%$ of light in the optical regime when the radiation rate of the guided mode in the photonic crystal is equal to the absorption rate in graphene\cite{Piper2014, Piper2014a}. Afterwards, this mechanism inspired various designs of photonic crystal integrating with the whole family of 2D materials, including graphene\cite{Lu2015, Guo2016, Jiang2017, Akhavan2018, Wang2019, Xiao2020a}, transitional metal dichalcogenides\cite{Huang2016, Li2017, Jiang2018, Hong2019, Cao2020, Wang2020}, black phosphorus\cite{Qing2018, Xiao2019, Liu2019, Liu2020}, and halide perovskites\cite{Cheng2020}. Actually, the radiation engineering is in principle a general way for the light absorption controlling, not limited to a specific mode in a certain resonator, say, guided mode in photonic crystal, but also applicable to other resonance modes in seemly different structures. With this consideration, the potential of dielectric metasurfaces for critical coupling is highly desirable in that they can provide unprecedentedly diverse arrangements and thus achieve full-control of light absorption.

In this work, we show a close relationship between the radiation engineering at critical coupling and the symmetry-protected BIC. Based on this, we propose a critical coupling system composed of silicon nanodisk metasurfaces supporting magnetic dipole quasi-BIC resonance and graphene. When the radiation rate of the resonance mode is equal to the dissipation loss of graphene, the maximum absorption of graphene is achieved under critical coupling conditions. Furthermore, the absorption bandwidth can be adjusted more than two orders of magnitude by changing the structure parameters of nanodisk, the Fermi level and the layer number of graphene. This work demonstrates the key role of BIC physics in the radiation engineering and set an example of designing novel critical coupling system based on dielectric metasurfaces, which hold great potentials for the next-generation 2D material optical and photonic devices.

\section{\label{sec2}Theoretical derivations}

Based on the temporal coupled-mode theory (CMT), we consider a single-mode optical resonator with amplitude $a$ coupled with two identical ports. For this system, the dynamic equations can be written as\cite{Haus1984, Fan2003},
\begin{eqnarray}
\frac{da}{dt}&=&(i\omega_{0}-\gamma-\delta)a+D^{T}|s_{+}\rangle,\label{eq1} \\
|s_{-}\rangle&=&C|s_{+}\rangle+Da,\label{eq2}
\end{eqnarray}
where $\omega_0$ is the resonance frequency, $\gamma$ and $\delta$ represent the radiation and the dissipative loss rates, respectively. $|s_{+}\rangle=[s_{1+},s_{2+}]^{T}$ and $|s_{-}\rangle=[s_{1-},s_{2-}]^{T}$ are amplitudes of incoming and outgoing waves. $C$ is the scattering matrix of the direct (non-resonant) process, which describes the transmission and reflection between the ports in the absence of the resonator, whereas $D$ is the coupling matrix account for coupling between each ports and resonance state $a$, which takes the form $D=[d_{1},d_{2}]^{T}$ with $D^{\dagger}D=2\gamma$ and $CD^{\ast}=-D$, due to the electromagnetic reciprocity and energy conservation. With the incoming wave amplitude $|s_{+}\rangle$, the amplitude of the resonance mode is,
\begin{equation}
a=\frac{D^{T}|s_{+}\rangle}{i(\omega-\omega_{0})+(\gamma+\delta)}.\label{eq3}
\end{equation}
The light absorption at the frequency $\omega$ can be expressed as,
\begin{equation}
\begin{split}
A
&= \frac{\langle s_{+}|s_{+}\rangle-\langle s_{-}|s_{-}\rangle}{\langle s_{+}|s_{+}\rangle}=\frac{2\delta|a|^{2}}{\langle s_{+}|s_{+}\rangle}      \\
&= \frac{2\delta\gamma}{(\omega-\omega_{0})^{2}+(\gamma+\delta)^{2}}.\label{eq4}
\end{split}
\end{equation}
It is easy to find that the absorption spectrum satisfies the Breit-Wigner distribution, and this formula predicts a symmetric Lorentzian curve which characterized by three parameters, $\omega_{0}$, $\gamma$, and $\delta$. It can be seen that the absorption reach its maximum at the resonance frequency $\omega=\omega_{0}$, and the peak $A_0$ is determined by the ratio between $\gamma$ and $\delta$,
\begin{eqnarray}
A_{0}=\frac{2}{\frac{\gamma}{\delta}+\frac{\delta}{\gamma}+2}.\label{eq5}
\end{eqnarray}
When the critical coupling condition $\gamma=\delta$ is satisfied, there is a theoretically maximum $A_{0}=0.5$ for this single-mode two-port system. Moreover, the absorption bandwidth defined as the full width at half maximum (FWHM) $\Gamma^{\text{FWHM}}=2|\omega_{1}-\omega_{0}|$ with $A_{1}=A_{0}/2$ at $\omega_{1}$ can be derived,
\begin{eqnarray}
\Gamma^{\text{FWHM}}=2(\gamma+\delta)=4\gamma.\label{eq6}
\end{eqnarray}

In order to get full control of the light absorption, the radiation rate $\gamma$, i.e., the inverse radiation lifetime, should be meticulously engineered according to the deep physical mechanism of BIC. For photonic structures satisfy the in-plane symmetry, the symmetry-protected BIC exists in the momentum space, which means the radiation channel is totally forbidden, and it can be treated as a resonance with infinite lifetime and vanishing linewidth $\gamma=0$. As the geometric parameters evolve, the in-plane symmetry is broken due to the perturbation, the BIC will turn to the quasi-BIC with infinitesimal inverse radiation lifetime $\gamma>0$ and a radiation channel with the background field is opened, generating a leak mode with asymmetric Fano line shape. To describe this sort of situation, $\gamma$ can be calculated analytically by the sum of radiations from each ports and incoming and outgoing waves\cite{Koshelev2018, Li2019},
\begin{eqnarray}
\gamma=c\sum_{i=x,y}|D_{i}|^{2},\label{eq7}
\end{eqnarray}
with the coupling amplitude $D_i$,
\begin{eqnarray}
D_{x,y}=-\frac{\omega_{0}}{\sqrt{2S_{0}}c}(p_{x,y}\mp\frac{1}{c}m_{y,x}+\frac{ik_{0}}{6}Q_{xz,yz}),\label{eq8}
\end{eqnarray}
where $S_{0}$ is the area of unit lattice, $p$, $m$, and $Q$ are the components of electric dipole, magnetic dipole, and electric quadrupole moments in the irreducible representations, respectively. Considering a situation where the in-plane symmetry along the $y$ axis is broken, it follows that $D_{x}=0$ since the electric filed components $E_{y}$ and $E_{x}$ are classified as even and odd functions with respect to symmetry. Moreover, $m_{x}$ and $Q_{yz}$ are both equal to zero because of the up-down mirror symmetry, $E_{y}(-z)=E_{y}(z)$. Then a simplified expression is obtained,
\begin{eqnarray}
\gamma=\frac{\omega_{0}^{2}}{2S_{0}c}\left|p_{y}\right|^2.\label{eq9}
\end{eqnarray}
$p_{y}=\pm \alpha p_{0}$ is defined as the net dipole moment in an asymmetric resonator, where $p_{0}$ is the electric dipole moment in the corresponding symmetry case and $\alpha$ is the so-called asymmetry parameter. The radiation rate of quasi-BIC can be rewritten as a function of asymmetry parameter,
\begin{eqnarray}
\gamma=\frac{\omega_{0}^{2}}{2S_{0}c}|p_{0}|^{2}\alpha^{2}.\label{eq10}
\end{eqnarray}
When the critical coupling condition is satisfied, the absorption bandwidth $\Gamma^{\text{FWHM}}$ can be reached by a simple expression via substituting Eq. (\ref{eq10}) to (\ref{eq6}),
\begin{eqnarray}
\Gamma^{\text{FWHM}}=\frac{2\omega_{0}^{2}}{S_{0}c}|p_{0}|^{2}\alpha^{2}.\label{eq11}
\end{eqnarray}
It can be found that the absorption bandwidth at critical coupling is proportional to the square of the asymmetric parameter, i.e., $\Gamma^{\text{FWHM}}\propto\alpha^{2}$, and all the variables in Eq. (\ref{eq11}) is decided by the geometry parameters of photonic structure. Following this way, the light absorption at critical coupling can be precisely controlled in quasi-BIC metasurface systems.

\section{\label{sec3}Numerical results and discussions}

\begin{figure}[htbp]
\centering
\includegraphics
[scale=0.40]{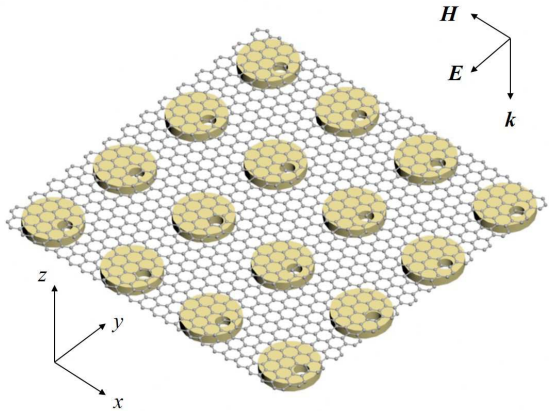}
\caption{\label{fig1} Schematic of the proposed critical coupling system composed of graphene on top of silicon nanodisk metasurfaces.}
\end{figure}

Here we consider a typical example where 2D material graphene is deposited on top of nanodisk metasurfaces supporting magnetic dipole quasi-BIC resonance, as shown in Fig. \ref{fig1}. The nanodisk itself has mirror symmetry in the out-of-plane direction. The graphene adds dissipative loss to the system, but it is so thin that it hardly breaks the up-down symmetry. The unit cell is periodically arranged with a lattice constant $p=1000$ nm, and the radius and thickness of the nanodisk are $R=300$ nm and $h=100$ nm, respectively. An off-centered hole with a variable radius $r$ is introduced with a fixed distance $d=150 $ nm away from the center of the nanodisk. The in-plane symmetry is thus broken, which allows for the establishment of radiation channel and transition of the resonance state from the symmetric-protected BIC into the quasi-BIC.    

\begin{figure}[htbp]
\centering
\includegraphics
[scale=0.45]{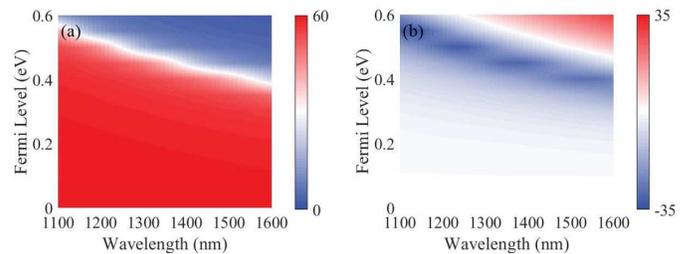}
\caption{\label{fig2} (a) Real part and (b) imaginary part of optical conductivity of graphene as a function of wavelength and Fermi level.}
\end{figure}

Due to the high refractive index and negligible absorption in the study range of near-infrared, silicon is chosen to be the construction material of nanodisk metasurfaces. For simplicity, a real constant $n_{\text{Si}}=3.5$ is adopted. The optical conductivity of graphene is derived from the random phase approximation (RPA) in the local limit including intraband and interband contributions\cite{Zhang2015, Xiao2016}, 
\begin{eqnarray}
\sigma_{\text{g}} &=&\frac{2e^{2}k_{\text{B}}T}{\pi\hbar^2}\frac{i}{\omega+i\tau^{-1}}\ln\bigg[2\cosh\bigg(\frac{E_{\text{F}}}{2k_{\text{B}}T}\bigg)\bigg]  \nonumber\\
&+&\frac{e^{2}}{4\hbar}\bigg[\frac{1}{2}+\frac{1}{\pi}\arctan\bigg(\frac{\hbar\omega-2E_{\text{F}}}{2k_{\text{B}}T}\bigg)  \nonumber\\
&-&\frac{i}{2\pi}\ln\frac{(\hbar\omega+2E_{\text{F}})^{2}}{(\hbar\omega-2E_{\text{F}})^{2}+4(k_{\text{B}}T)^{2}}\bigg],\label{eq12}
\end{eqnarray}
where $e$, $k_{B}$, $T$, $\hbar$, and $\omega$ are the electron charge, the Boltzmann constant, the room temperature, the reduced Planck’s constant, and the incident light frequency, respectively. $\tau$ is the relaxation time and depends on the carrier mobility $\mu$, the Fermi level $E_{\text{F}}$ and the Fermi velocity $v_{\text{F}}$ with the relation $\tau=(\mu E_{\text{F}})/(e v_{\text{F}}^{2})$, where moderate measured values $\mu=10000$ cm$^{2}$/V$\cdot$s and $v_{\text{F}}=1\times 10^{6}$ m/s are adopted. The optical conductivity of graphene can be efficiently controlled by varying the Fermi level. As shown in Figs. \ref{fig2}(a) and \ref{fig2}(b), the real part of optical conductivity reduces obviously once the Fermi level increases from the Dirac point by half photon energy due to the Pauli blocking in the near-infrared, while the imaginary part continuously increases. The tunable optical conductivity of graphene lays the foundation for the control of the dissipative loss rate in the critical coupling system. 

The full-wave numerical simulations are performed using the
finite-difference time-domain method (FDTD Solutions, Lumerical
Inc., Canada). In the following calculations, a mesh grid $20\times20\times20$ nm together with an auto shutoff min $1e-07$ is adopted to make good trade-off between accuracy, RAM capacity and running time. The convergence test results including the maximum field intensity and the estimated running time are attached in the Supplemental Material\cite{Xiao2020SM}. The $y$-polarized plane wave is incident along the -$z$ direction, and accordingly, periodic boundary conditions are utilized in the $x$-$y$ plane and the perfect matching layers are adopted in the $z$ direction.

\begin{figure}[htbp]
\centering
\includegraphics
[scale=0.45]{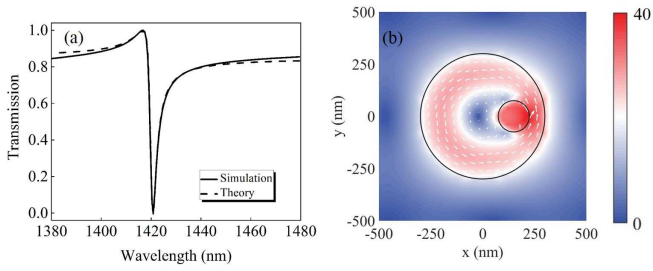}
\caption{\label{fig3} (a) Simulated and fitted transmission spectra of nanodisk metasurfaces. (b) Magnitude of the electric field across the unit cell at the resonance, overlaid with arrows indicating the direction of displacement current.}
\end{figure}

To understand the radiation engineering through metasurface governed by BIC, we first perform resonance mode analyses without the presence of graphene. At a radius of the off-centered hole $r=75$ nm, the quasi-BIC obtains an energy exchange with the continuum free-space radiation modes, and manifest itself as a sharp Fano resonance. As shown in Fig. \ref{fig3}(a), The transmission spectrum exhibits an asymmetric line shape with a narrow dip at $1420.91$ nm, which is well fitted by the classical Fano formula within the CMT framework\cite{Hsu2013, Wu2019}. The radiation rate of the resonance mode can be extracted as $\gamma=1.315$ THz, while the dissipative loss rate is zero, since silicon is considered as lossless material here. The corresponding near field distribution in the $x$-$y$ plane is shown in Fig. \ref{fig3}(b). The maximum enhancement of the electric field is located at the center of the off-centered hole, and the overlaid arrows indicate the circulation distribution of the displacement current, which reveal that the incident light is strongly trapped within the metasurface by the magnetic dipole moment oscillation along the $z$ direction. To further confirm the magnetic dipole quasi-BIC resonance, the contributions of multipole moments to the far-field radiation are decomposed under the Cartesian coordinate system. As shown in the Supplemental Material, the radiation power of the magnetic dipole moment (MD) is at least an order of magnitude greater than that of other multipole moments at the resonance, which reveals out the dominating role of the magnetic dipole, and in return, validates the single-mode assumption in the theoretical derivations\cite{Xiao2020SM}.

\begin{figure}[htbp]
\centering
\includegraphics
[scale=0.45]{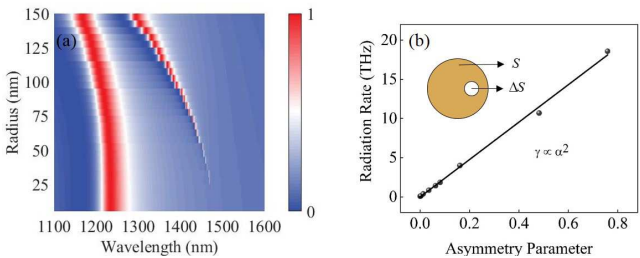}
\caption{\label{fig4} (a) Transmission spectra of nanodisk metasurfaces as a function of the radius of the off-centered hole. (b) Values of the radiation rate $\gamma$ as a function of the asymmetry parameter $\alpha$. Inset: the asymmetry parameter is defined as $\alpha=\Delta S/S$, where $\Delta S$ and $S$ are the areas of the off-centered hole and the nanodisk, respectively.}
\end{figure}

The quasi-BIC resonance is closely related to the radiation leakage, hence the radiation rate and the transmission bandwidth can be adjusted through manipulating the geometry parameters. With the evolution of the radius of the off-centered hole, the transmission spectra show an increasing bandwidth at the magnetic dipole quasi-BIC resonance, as shown in Fig. \ref{fig4}(a). The bandwidth expansion can be attributed to the increase in the radiation rate with the asymmetry degree. In a quantitative description, here we define the asymmetric parameter as the ratio of the area of the hole to that of the nanodisk, i.e., $\alpha=\Delta S/S$. The values of radiation rate $\gamma$ as a function of the asymmetry parameter $\alpha$ for the proposed nanodisk metasurfaces are summarized in Fig. \ref{fig4}(b). A quadratic dependence of $\gamma$ on $\alpha$ can be nicely observed, which is consistent with the theoretical derivation in Eq. (\ref{eq10}) for the quasi-BIC resonance in asymmetric metasurfaces. To investigate the array effects in the magnetic dipole quasi-BIC resonance, the responses of the finite sized arrays of resonators, i.e., the individual nanodisk, $3\times3$, $5\times5$, $7\times7$, and $9\times9$ nanodisk arrays, are considered in the Supplemental Material\cite{Xiao2020SM}. The required size of the nanodisk array for a robust magnetic dipole quasi-BIC resonance is expected not large.

\begin{figure}[htbp]
\centering
\includegraphics
[scale=0.45]{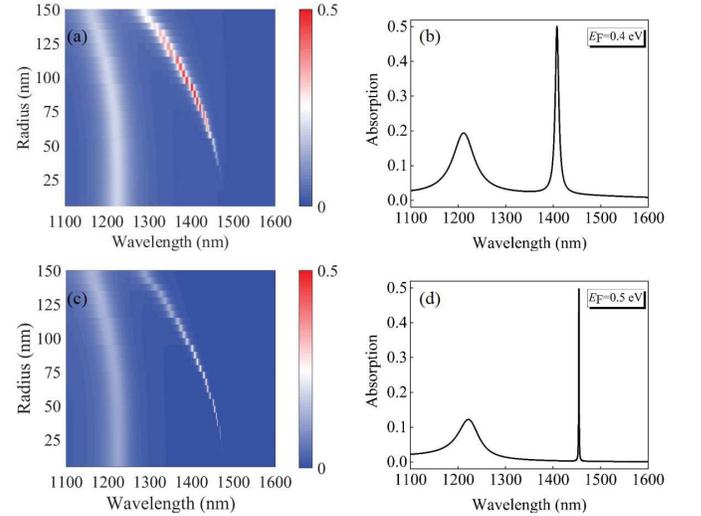}
\caption{\label{fig5} Realization of narrowband absorption of graphene by shifting the Fermi level. (a) and (c) show the absorption spectra as a function of wavelength and radius of the off-centered hole, and (b) and (d) provide the corresponding absorption spectra at critical coupling for monolayer graphene with different Fermi levels of $E_{\text{F}}=0.4$ and $0.5$ eV.}
\end{figure}

To construct a critical coupling system, a monolayer graphene is introduced as a lossy component on top of nanodisk metasurfaces, which determines the dissipative loss rate of the system. When the radiation rate of the quasi-BIC resonance equals the dissipative loss rate of graphene, i.e., $\gamma=\delta$, the critical coupling condition is satisfied and the maximum absorption of $0.5$ is achieved. We first focus on the realization of narrowband absorption at critical coupling when graphene becomes doped. As mentioned above, the real part of optical conductivity of graphene shows a steplike decline if the Fermi level increases from the Dirac point by half photon energy, i.e., $E_{\text{F}}>\hbar\omega/2$, which will remarkably reduce the dissipative loss. To this end, the Fermi levels of $0.4$ eV and $0.5$ eV are considered in the wavelength of interest. The absorption spectra are plotted as a function of wavelength and radius of the off-centered hole, as shown in Figs. \ref{fig5}(a) and \ref{fig5}(c), the white contour line accurately draws the outline of the FWHM. It can be seen that the absorption bandwidth rapidly declines to an ultra-narrow profile when the Fermi level increases from $0.4$ to $0.5$ eV. The corresponding critical coupling points can be found by sweeping the wavelength and radius. As shown in Figs. \ref{fig5}(b) and \ref{fig5}(d), the maximum absorption is achieved at 1408 nm with $r=85$ nm for $E_{\text{F}}=0.4$ eV and at 1454.48 nm with $r=45$ nm for $E_{\text{F}}=0.5$ eV, respectively. The values of the radiation rate and dissipative loss rate can be extracted as $\gamma=\delta=2.254$ and $0.201$ THz. As a result of the simultaneous decreasing of $\gamma$ and $\delta$, the absorption bandwidth shrinks by more than one order of magnitude from $9.02$ nm to $0.9$ nm.

\begin{figure}[htbp]
\centering
\includegraphics
[scale=0.65]{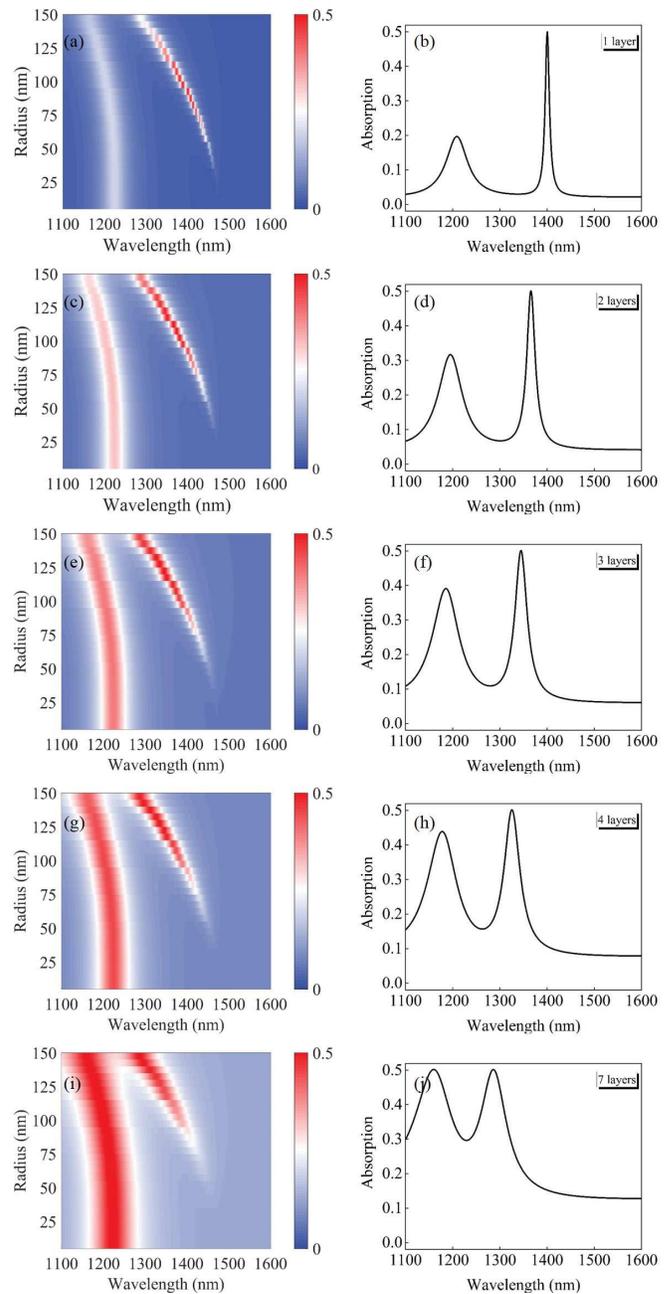}
\caption{\label{fig6} Realization of broadband absorption of graphene by adding the layer number. (a), (c), (e), (g), and (i) show the absorption spectra as a function of wavelength and radius of the off-centered hole, and (b), (d), (f), (h), and (j) provide the corresponding absorption spectra at critical coupling for multilayer graphene with different layer numbers of $N=1$, $2$, $3$, $4$, and $7$.}
\end{figure}

The above discusses the realization of narrowband absorption of graphene by shifting the Fermi level. On the other hand, the possibility of broadband absorption can also be explored by simultaneously increasing the radiation rate of the quasi-BIC magnetic dipole resonance and the dissipate loss rate of graphene, which may have potential in broadband applications, such as photodetection, light energy harvesting, and so on. To maximize the dissipate loss, the monolayer graphene deposited on top of silicon nanodisk metasurfaces is replaced with multilayer graphene at the charge neutral point, i.e., $E_{\text{F}}=0$ eV. The optical conductivity shows a linear increase with the layer number\cite{Hass2008, Li2016}. As shown in Figs. \ref{fig6}(a), \ref{fig6}(c), \ref{fig6}(e), \ref{fig6}(g), and \ref{fig6}(i), the absorption spectra are plotted as a function of wavelength and radius of the off-centered hole, in which the absorption bandwidth increase continuously when the layer number increases from $N=1$ to $7$. The corresponding critical coupling points can be found by sweeping the wavelength and radius. As shown in Figs. \ref{fig6}(b), \ref{fig6}(d), \ref{fig6}(f), \ref{fig6}(h), and \ref{fig6}(j), the maximum absorption is achieved at 1400.54 nm with $r=90$ nm for $N=1$, at 1365.73 nm with $r=110$ nm for $N=2$, at 1344.84 nm with $r=120$ nm for $N=3$, at 1325.75 nm with $r=130$ nm for $N=4$, and at 1285.83 nm with $r=145$ nm for $N=7$, respectively. The values of the radiation rate and dissipative loss rate can be extracted as $\gamma=\delta=2.712$, $7.44$, $9.078$, $12.505$, and $21.375$ THz. As a result of the simultaneous increasing of $\gamma$ and $\delta$, the absorption bandwidth expands by nearly one order of magnitude from $10.87$ nm to $94.53$ nm.

\begin{figure}[htbp]
\centering
\includegraphics
[scale=0.40]{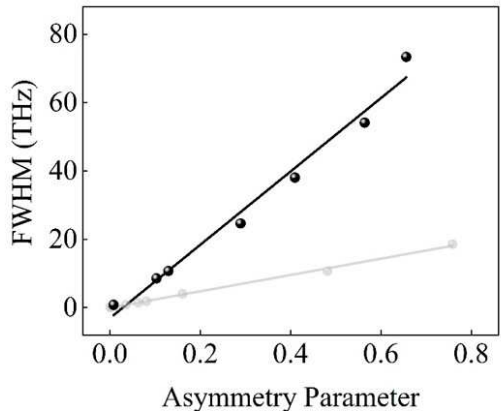}
\caption{\label{fig7} Quadratic dependence of the absorption bandwidth $\Gamma^{\text{FWHM}}$ (black line) and the radiation rate $\gamma$ (grey line) on the asymmetry parameter $\alpha$.}
\end{figure}

Finally, we examine the dependence of the absorption bandwidth on the asymmetry parameter. The values of $\Gamma^{\text{FWHM}}$ are estimated from the absorption spectra of Figs. \ref{fig5} and \ref{fig6}, corresponding to the cases with different Fermi levels and layer numbers of graphene. As shown in Fig. \ref{fig7}, $\Gamma^{\text{FWHM}}$ is proportion to the square of $\alpha$ (black line), i.e., $\Gamma^{\text{FWHM}}\propto\alpha^{2}$, and in particularly, the slope of such linear fitting is approximately four times that of the radiation rate $\gamma$ (grey line), which verifies the theoretical prediction in Eq. (\ref{eq11}). It should be noted that the quadratic dependence of the radiation rate $\gamma$ on the asymmetric parameter $\alpha$ holds for the Fano resonance governed by the BIC physics in asymmetry metasurfaces, which is not limited to the magnetic dipole resonance, but also goes for other non-radiative states such as toroidal mode, anapole mode, and supercavity mode. Therefore, the quadratic scalability of $\Gamma^{\text{FWHM}}$ at critical coupling together with the light absorption controlling strategy shows general applicability.

\section{\label{sec4}Conclusions}

In summary, we develop a general method to control light absorption at critical coupling through the quasi-BIC resonance, and demonstrate it in a typical single-mode two-port system composed of graphene coupled with silicon nanodisk metasurfaces. When the radiation rate of the magnetic dipole resonance is equal to the dissipate loss rate of graphene, the critical coupling condition is satisfied and the maximum absorption of 0.5 is obtained. By simultaneously changing the asymmetric parameter of metasurfaces, the Fermi level and the layer number of graphene, the absorption bandwidth can be flexibly adjusted more than two orders of magnitude from $0.9$ nm to $94$ nm in the near-infrared. More interesting, the quadratic dependence of the absorption bandwidth on the asymmetry parameter is predicted and verified, which reveals out the key role of BIC physics in the radiation engineering. Beyond the proposed example, the theoretical method can be applied to a various types of critical coupling systems, in principle, where different atomically thin 2D materials coupled with asymmetric metasurfaces supporting quasi-BIC resonance. With this consideration, this work not only gains a deeper physical insight into the metasurface-mediated BIC physics, but also paves a way towards smart design of 2D material devices such as light emitters, detectors, modulators, and sensors.

\begin{acknowledgments}	
This work is supported by the National Natural Science Foundation of China (Grants No. 61775064, No. 11847132, No. 11947065, No. 61901164, and No. 12004084), the Natural Science Foundation of Jiangxi Province (Grant No. 20202BAB211007), the Interdisciplinary Innovation Fund of Nanchang University (Grant No. 2019-9166-27060003), the Natural Science Research Project of Guizhou Minzu University (Grant No. GZMU[2019]YB22), and the China Scholarship Council (Grant No. 202008420045). The authors would also like to thank Dr. S. Li for her guidance on the effective multipole expansion and Dr. X. Jiang for beneficial discussions on the critical coupling mechanism.

X.W. and J.D. contributed equally to this work.
\end{acknowledgments}

%

\end{document}